\documentclass[pdflatex]{ptephy_v1}

\preprintnumber{XXXX-XXXX}

\usepackage{graphicx}
\usepackage{amsmath}
\usepackage{here}
\usepackage{mathtools}
\usepackage{enumerate}

\usepackage{color}

\begin{document}

\title{Wide angle acceptance and high-speed track recognition in nuclear emulsion}

\author[1]{Y.~Suzuki}
\affil{Graduate School of Science, Nagoya University, Nagoya 464-8602, Japan}

\author[1,*]{T.~Fukuda}

\author[1]{H.~Kawahara}

\author[1]{R.~Komatani}

\author[1]{M.~Naiki}

\author[1,2]{T.~Nakano}
\affil {Kobayashi-Maskawa Institute for the Origin of Particles and the Universe, Nagoya University, Nagoya 464-8602, Japan}

\author[3]{T.~Odagawa}
\affil{Department of Physics, Kyoto University, Kyoto 606-8502, Japan}

\author[4]{M.~Yoshimoto}
\affil{RIKEN Nishina Center for Accelerator-Based Science, RIKEN, Saitama 351-0198, Japan\email{tfukuda@flab.phys.nagoya-u.ac.jp}}

\begin{abstract}A nuclear emulsion film is a three-dimensional tracking device that is widely used in cosmic-ray and high energy physics experiments.
Scanning with a wide angle acceptance is crucial for obtaining track information in emulsion films.
This study presents a new method developed for wide angle acceptance and high-speed track recognition of nuclear emulsion films for neutrino-nucleus interaction measurements.
The nuclear emulsion technique can be used to measure tracks of charged particles from neutrino interactions with a low momentum threshold.
The detection of the particles with a wide angle acceptance is essential for obtaining detailed information on the interactions in the sub- and multi-GeV neutrino energy region.
In the new method developed for a neutrino interaction measurement in J-PARC called NINJA, the angle acceptance is covered up to $|\tan\theta_{x(y)}| < 5.0$ (80\% of all solid angles) with $150\,\mathrm{m^2/year}$.
This method can also be used to improve the angle accuracy and recognition efficiency of the tracks.
 \end{abstract}

\subjectindex{H16, H34}

\maketitle

\section{Introduction\label{sec:introduction}}

A nuclear emulsion is a three-dimensional tracking device which is widely used in cosmic-ray and high energy physics experiments.
It has been crucial in discoveries of pions~\cite{Lattes:1947mw} and charmed particles~\cite{Niu:1971xu} in cosmic ray, the direct detection of $\nu_\tau$ interactions in the DONUT experiment~\cite{DONUT:2000fbd}, and the direct detection of the $\nu_\mu \to \nu_\tau$ appearance in the OPERA experiment~\cite{OPERA:2015wbl} owing to its high granularity.

The films must be developed after exposure to analyze the tracks recorded in the emulsion films.
Then the tracks can be observed through optical microscopes.
Automatic scanning and track recognition are necessary to measure a large number of tracks in the entire area of the films.
The automatic scanning system, so called Track Selector, was first developed in Nagoya University and was subsequently applied in several experiments~\cite{Niwa:1974, Aoki:1989uk}.
Ultra Track Selector was developed in 1998 and was used in the DONUT and CHORUS experiments~\cite{Nakano:dt}.
Super-Ultra Track Selector (S-UTS) was then developed in 2010~\cite{Morishima:2010zz} with a scanning speed of $\sim 10\,\mathrm{m^2}/\mathrm{year}$ and the angle acceptance of $|\tan\theta| < 0.5$, for the OPERA experiment.
Here, $\theta$ is an angle recognized by the scanning system with respect to the direction perpendicular to the film surface.
The requirement for the speed has been increasing in recent years following the success of S-UTS in the OPERA experiment, and a new generation scanning system, called Hyper Track Selector (HTS) has been developed and is currently being operated in Nagoya University with a scanning speed of $\sim 1000\,\mathrm{m^2}/\mathrm{year}$~\cite{Yoshimoto:2017ufm} since 2014.

The nuclear emulsion itself has an acceptance of $4\pi$.
Scanning with a wide angle acceptance is crucial for obtaining track information in emulsion films.
Fine Track Selector (FTS) was developed in the OPERA experiment for identification of hadron interaction background from signal tau decays by detecting nuclear fragment with wide angle acceptance in 2012~\cite{Fukuda:2013nq,Ishida:2014qga}.
In FTS, it was found that very high detection efficiency could be obtained even for the large angle minimum ionizing particle tracks~\cite{Fukuda:2014vda}.
The angle acceptance of the FTS was $|\tan\theta| < 3.5$ but the speed was only sufficient to scan around the interaction points.

This result opened up a detailed study of low energy neutrino interactions with many large angle particles using nuclear emulsion detectors.
The conventional HTS scanning method specializing in scanning speed presented an angle acceptance of $|\tan\theta| < 1.5$ for all areas of the films.

A neutrino interaction measurement using nuclear emulsion films has been implemented in Japan Proton Accelerator Research Complex (J\nobreakdash-PARC)~\cite{Fukuda:2017clt, Yamada:2017qeg, Hiramoto:2020gup, Oshima:2020ozn} since 2014.
The Neutrino Interaction research with Nuclear emulsion and J\nobreakdash-PARC Accelerator (NINJA) experiment aims to measure the neutrino\nobreakdash-nucleus interactions in the sub- and multi-GeV neutrino energy region with a high-intensity neutrino beam from the J\nobreakdash-PARC neutrino beamline.
The uncertainty of the neutrino interactions in this region is one of the main sources of the systematic uncertainties in the current and future long-baseline neutrino oscillation experiments~\cite{Abe:2019vii, NOvA:2021nfi, Abe:2018uyc, Acciarri:2015uup}.
The particles, mainly protons or pions, tend to be emitted with a large angle with respect to the neutrino beam direction in this energy region\footnote{Mean neutrino energies of the DONUT or OPERA experiment are $\mathcal{O}(10\text{--}100)\,\mathrm{GeV}/c$.}.
Therefore, a new method of recognizing large angle tracks of up to $|\tan\theta| < 5.0$ using HTS with a sufficient speed to scan the entire area of the emulsion films is developed.

In this paper, we present a new method of track reconstruction with a wide angle acceptance and at high-speed, and demonstrate its performance.
Sect.~\ref{sec:detector} presents the detector setup and the conditions of the tracks used in the evaluation.
Sect.~\ref{sec:analysis} explains the methods of track recognition in the emulsion gel and films, and also explains the noise rejection.
Sect.~\ref{sec:performance} presents the evaluation of the recognition efficiency, angle accuracy of the recognized tracks, and the processing speed of the method and Sect.~\ref{sec:conclusion} concludes the paper.
 
\section{Detectors used for performance evaluation\label{sec:detector}}

\begin{figure}[H]
    \centering 
    \includegraphics[width = 0.75\textwidth]{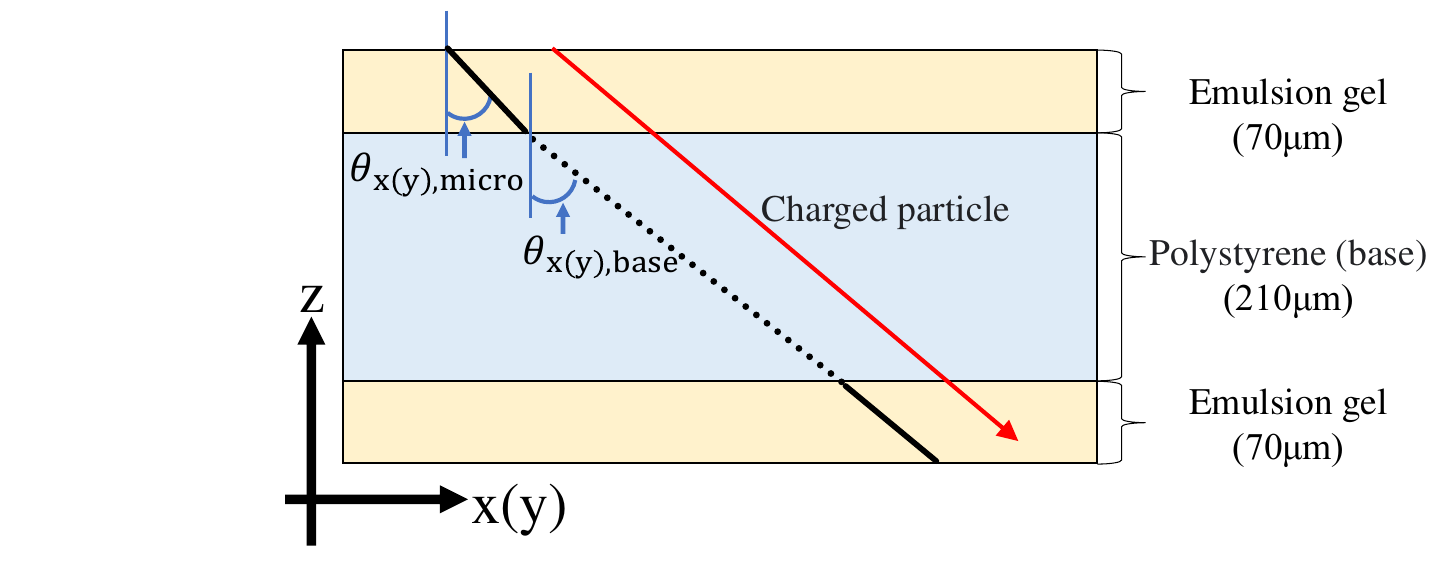}
    \caption[Film cross section]{Cross-sectional view of an emulsion film used in the NINJA experiment. Two emulsion gel layers are applied on both sides of a polystyrene layer. The black solid lines represent microtracks and the dotted line represents a basetrack. The direction perpendicular to the film surface is defined as the $z$ direction, and the $x$ and $y$ directions exist on the film surface. $\theta_{x(y),\mathrm{micro}}$ is an angle of a microtrack on the $x(y)\text{-}z$ plane and $\theta_{x(y),\mathrm{base}}$ is an angle of a basetrack on the plane.\label{fig:02:film_crosssection}}
\end{figure}

The data obtained in the J\nobreakdash-PARC E71a of the NINJA experiment are used to evaluate the new method of track recognition.
The emulsion films used in this measurement have two  emulsion gel layers with a thickness of $70\,\mathrm{\mu m}$ on both sides of a polystyrene sheet with a thickness of $210\,\mathrm{\mu m}$, as shown in Figure~\ref{fig:02:film_crosssection}.
The polystyrene sheet is called the ``base'' as it supports the gel layers.
The emulsion gel consists of silver halide crystals and gelatin with a volume ratio of $35:65$.
The direction perpendicular to the film surface is defined as the $z$ direction, and the $x$ and $y$ directions are on the film surface, as shown in Figure~\ref{fig:02:film_crosssection}, and are perpendicular to each other.

A ``microtrack'' is a track in the emulsion gel which is detected as silver grains linearly aligned in the emulsion gel.
A virtual track in the base is reconstructed as a ``basetrack'' when the microtracks detected in both layers of the emulsion gel are connected.
The angle accuracy of the basetracks is better than that of the microtracks since the base is more rigid and is thicker than the emulsion gel.
$\theta_{x(y),\mathrm{micro}}$ is an angle of a microtrack on the $x(y)\text{-}z$ plane and $\theta_{x(y),\mathrm{base}}$ is an angle of a basetrack on the plane with respect to the $z$ direction.

\begin{figure}[h]
    \centering 
    \includegraphics[width = 0.7\textwidth]{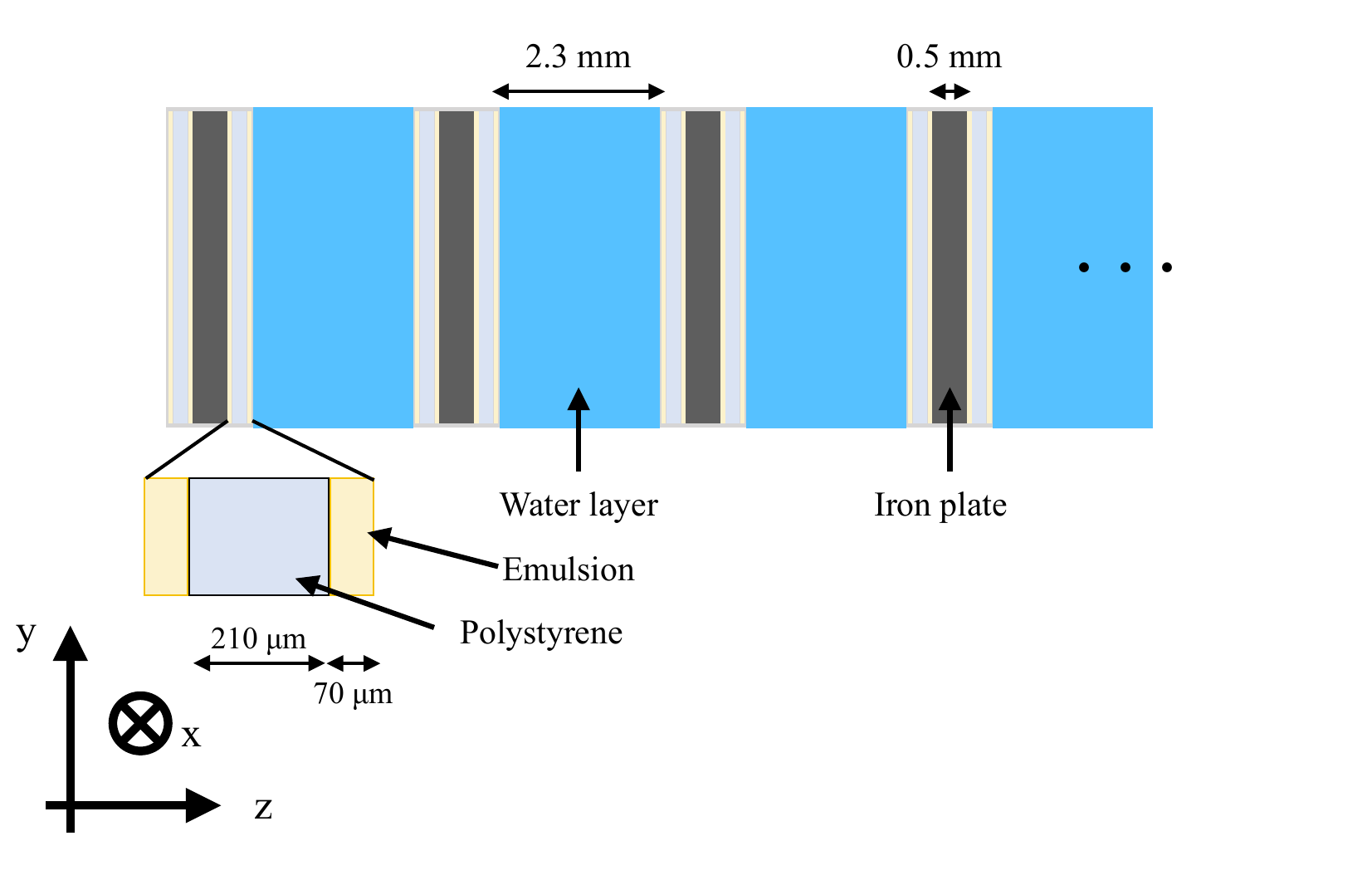}
    \caption[structure of ECC]{Structure of the Emulsion Cloud Chamber. The detector consists of an alternating structure of target material layers and tracking layers. The target material is water with a thickness of $2.3\,\mathrm{mm}$ and the tracking layer consists of vacuum-packed two emulsion films on both sides of a $500\,\mathrm{\mu m}$ thick iron plate. The $y$ direction is set to be vertical during the exposure.\label{fig:02:ecc_schematic}}
\end{figure}

\begin{figure}[H]
    \centering 
    \includegraphics[width = 0.6\textwidth]{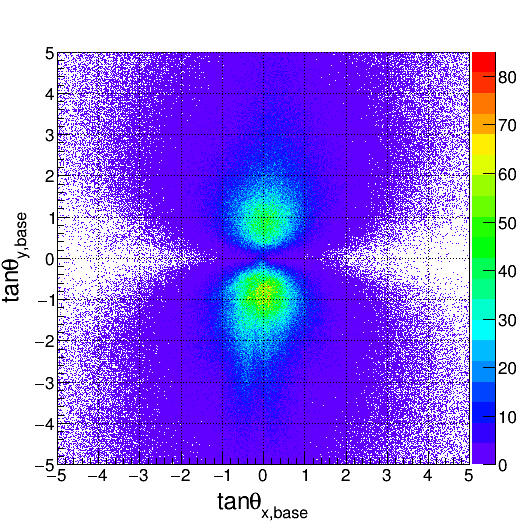}
    \caption[Angular distribution of track] {Angle distribution of the tracks in a single emulsion film. The tracks were accumulated during the beam exposure period. Only the tracks connected to the upstream film across one water layer are shown in this distribution.\label{fig:02:angle_distribution}}
\end{figure}

The sensitivity of the film is evaluated by the grain density, which is the number of grains along the $100\,\mathrm{\mu m}$ trajectory of the minimum ionizing particle.
The grain density of the particles recorded during the experimental period is measured to be $40\text{--}45\,\mathrm{grains/100\,\mu m}$~\cite{Takao:mt} and it is sufficiently high for the experiment.
The size of the film is $25\,\mathrm{cm} \times 25\,\mathrm{cm}$.
A single detector consists of emulsion films and materials, as shown in figure~\ref{fig:02:ecc_schematic}.
The detector is called an Emulsion Cloud Chamber (ECC) and consists of an alternate structure of tracking layers and $2.3\,\mathrm{mm}$ thick water layers.
A single tracking layer consists of vacuum-packed two emulsion films on both sides of a $500\,\mathrm{\mu m}$ thick iron plate.
The neutrino beam is almost parallel to the $z$ direction and the $y$ direction is vertical.
This ECC was placed $44\,\mathrm{m}$ underground in the J\nobreakdash-PARC Neutrino Monitor building.
The tracks of the cosmic muons were accumulated for approximately four months in the films of the ECC.
Figure~\ref{fig:02:angle_distribution} shows the angle distribution of the basetracks reconstructed in the ECC.
It shows a characteristic due to the zenith angle dependence of the cosmic muons and the acceptance of HTS.
The tracks in a single film are connected with the upstream adjacent film when the positions and angles satisfy certain connection allowances.
The track density during the construction of the ECC is approximately $5 \times 10^3\,\mathrm{tracks/cm^2}$ within $|\tan\theta_{x(y),\mathrm{base}}| < 5.0$.
The density of all tracks recorded in the film is estimated to be $1 \times 10^4\,\mathrm{tracks/cm^2}$.

\section{Track recognition algorithm\label{sec:analysis}}

\subsection{Conventional methods for microtrack recognition\label{ssec:analysis:microtrack_conventional}}
HTS reads out the tracks recorded in the emulsion gel~\cite{Yoshimoto:2017ufm}.
The size of a single view of HTS is $5.1\,\mathrm{mm} \times 5.1\,\mathrm{mm}$ and one view is taken by 72 sensors.
The size of each sensor is $0.92\,\mathrm{mm} \times 0.49\,\mathrm{mm}$, and it consists of $2048 \times 1088$ pixels, where the size of a single pixel is $0.45\,\mathrm{\mu m} \times 0.45\,\mathrm{\mu m}$ on the image plane.

The basic concept of the track recognition algorithms is identical to those presented in Refs.~\cite{Niwa:1974, Aoki:1989uk, Nakano:dt, Morishima:2010zz, Fukuda:2013nq, Yoshimoto:2017ufm,Fukuda:2014vda}.
In this algorithm, the scanning system captures 16 tomographic images for a single emulsion layer.
The pixels of the track are perpendicularly aligned with 16 relatively shifting tomographic images.
The track is detected with the angle corresponding to the shift amount when the total number of layers with a silver grain exceeds a threshold.
Therefore the angle acceptance can be expressed as $|\tan\theta_{x(y),\mathrm{micro}}| < D/L$, where $L$ is the thickness of 16 tomographic images and $D$ is the maximum shift.
Either a larger $D$ or a smaller $L$ are required to enlarge the angle acceptance of the track recognition.
The number of calculations is proportional to $D^2$ since the angle must be calculated in a two-dimensional angle space.
Thus, $L$ must be set to a smaller value to extend the angle acceptance with a limited calculation time.
However, the number of silver grains in a particle trajectory is proportional to $L$; thus, a smaller $L$ value results in lower track recognition efficiency.
Therefore, the angle acceptance is limited by the track recognition efficiency and the calculation time.

\subsection{New  methods for microtrack recognition\label{ssec:analysis:microtrack_new}}

\begin{figure}[h]
    \centering
    \includegraphics[width = 0.7\textwidth]{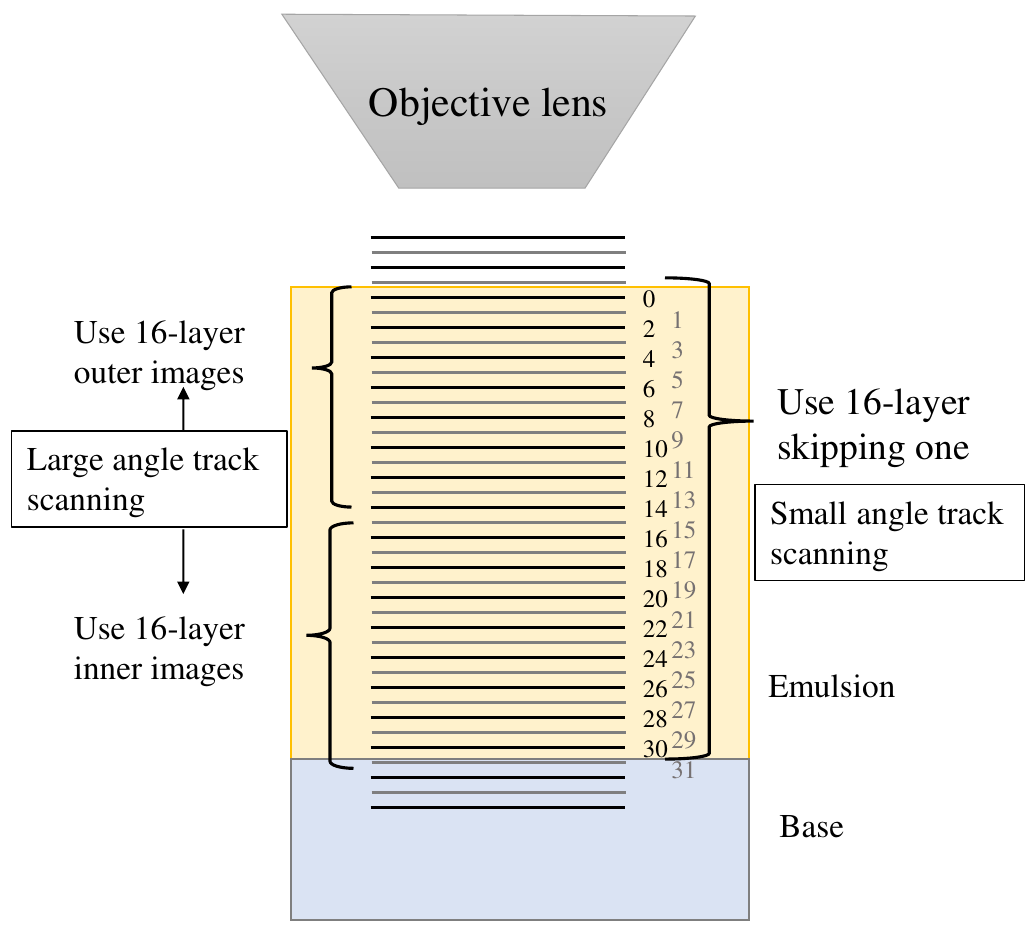}
    \caption[32 scan schematic]{Schematic view of 32 tomographic images for use of new scanning method. Small angle tracks are recognized using even (odd)-numbered tomographic images and large angle tracks are recognized using inner (outer) tomographic images.\label{fig:03:film_crosssection_32scan}}
\end{figure}

A new track recognition method is developed, which captures 32 tomographic images with a $2\,\mathrm{\mu m}$ pitch for a $62\,\mathrm{\mu m}$ emulsion gel layer and selects 16 appropriate ones to implement both the large angle acceptance and the high recognition efficiency.
The recognition procedure of the track in the emulsion gel, microtrack, is as follows:

\begin{enumerate}[\textbf{(}1\textbf{)}]
    \item \textbf{Image taking}\mbox{}\\
    HTS captures 32 tomographic images.
    Each tomographic image is binarized using the brightness values and the presence of the silver grain in each image is determined.
    Subsequently, each pixel containing the silver grain existence is expanded.
    One pixel is merged and expanded to $2 \times 2$ the original pixel size and the binarized images are stored in a storage server.
    \item \textbf{Track detection}\mbox{}\\
    Of the 32 images, 16 even (odd)-numbered or inner (outer) binarized ones are selected, and tracks (where the silver grains are linearly aligned) are searched for in these images, as shown in figure~\ref{fig:03:film_crosssection_32scan}.
    The position and angle of the tracks is recognized using the algorithm reported in Refs.~\cite{Niwa:1974, Aoki:1989uk, Nakano:dt, Morishima:2010zz, Fukuda:2013nq, Yoshimoto:2017ufm,Fukuda:2014vda}.
    \item \textbf{Track fitting}\mbox{}\\
    The position information of the pixels consisting of the trajectory is extracted from the 16 even-numbered binarized images using the position and angle of the recognized microtrack.
    Linear fitting is then applied to the pixels, after which the position and angle of the track is calculated.
\end{enumerate}

Figure~\ref{fig:03:film_crosssection_32scan} presents a schematic view of the new track recognition method.
Small angle tracks ($|\tan\theta_{x(y),\mathrm{micro}}| < 2.25$) are recognized using 16 even (odd)-numbered tomographic images which correspond to $L = 60\,\mathrm{\mu m}$ and $D = 135\,\mathrm{\mu m}$, and large angle tracks ($|\tan\theta_{x(y),\mathrm{micro}}| < 5.4$) are recognized using 16 inner (outer) images which correspond to $L = 30\,\mathrm{\mu m}$ and $D = 162\,\mathrm{\mu m}$.
It is also possible to recognize tracks with $D = 162\,\mathrm{\mu m}$ when $L = 60\,\mathrm{\mu m}$.
However, as for the tracks in the region of $|\tan\theta_{x(y),\mathrm{micro}}| > 2.0$, the track recognition efficiency is better with $L = 30\,\mathrm{\mu m}$.
To reduce the cost of the calculation, $D$ is set to $135\,\mathrm{\mu m}$ when $L = 60\,\mathrm{\mu m}$.
This method allows for the angle acceptance to be expanded while the accuracy of the small angle tracks is the same as in the conventional method.
The path length of the recognized track in the 16 tomographic images is expressed as $L \times \sqrt{1 + \tan^2\theta_{x,\mathrm{micro}} + \tan^2\theta_{y,\mathrm{micro}}}$ and the large angle tracks have longer paths, i.e., the number of silver grains of the large angle tracks is larger than that of the small angle tracks.
Therefore, the recognition efficiency of the large angle tracks does not decrease even for a small value of $L$.

The allowance of the track connection inside each emulsion film depends on the angle accuracy of the microtracks.
Here, the evaluation is applied to the tracks by using $|\tan\theta_{x(y),\mathrm{base}}| < 5.0$ which can be sufficiently connected.

\begin{figure}[h]
    \centering 
    \includegraphics[width = 0.75\textwidth]{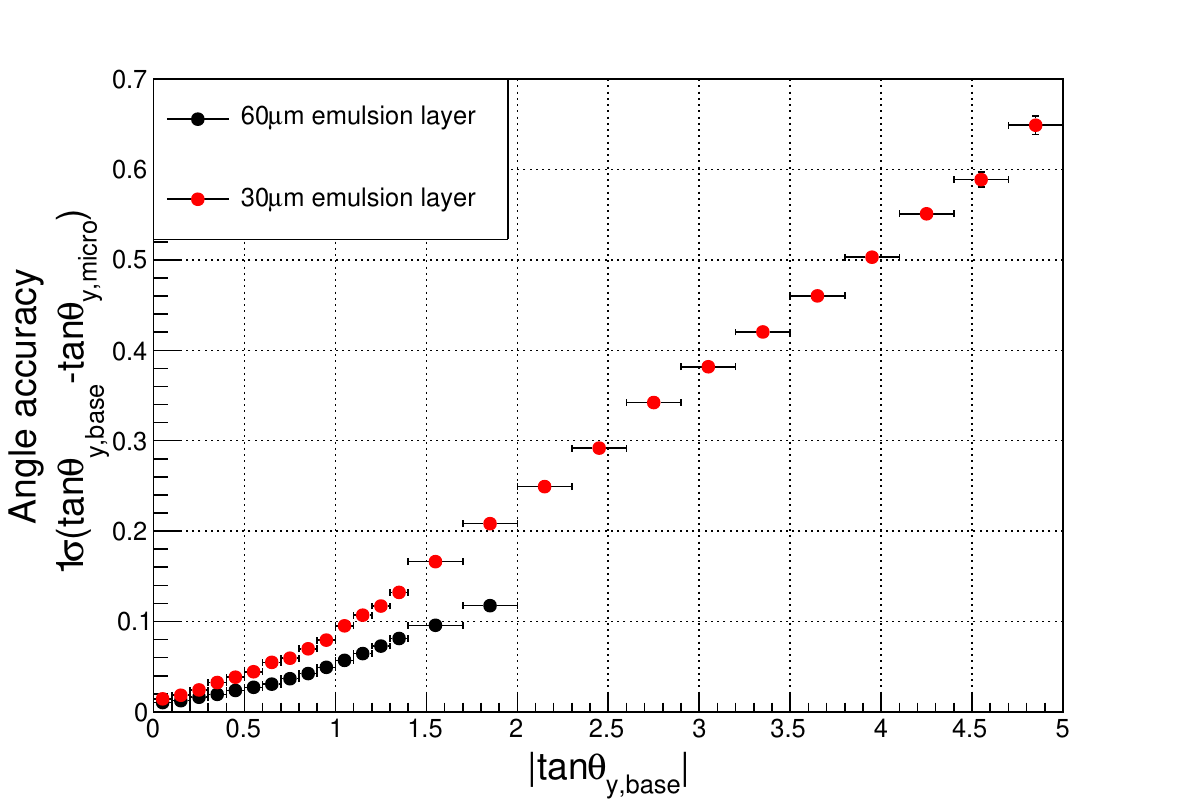}
    \caption[microtrack angle accuracy]{Angle accuracy of the microtracks. The accuracy is evaluated by the width of angular difference distributions between basetracks and microtracks. The black plot represents the accuracy of the microtracks recognized by the $60\,\mathrm{\mu m}$-thickness scanning, and the red plot represents the $30\,\mathrm{\mu m}$-thickness scanning. The angle acceptance of the $60\,\mathrm{\mu m}$-thickness scanning is limited to $|\tan\theta_{x(y),\mathrm{base}}| < 2.0$, whereas the $30\,\mathrm{\mu m}$-thickness scanning can extend up to $|\tan\theta_{x(y),\mathrm{base}}| < 5.0$.\label{fig:03:microtrack_angle_accuracy}}
\end{figure}

While the track recognition efficiency of the large angle tracks is expected to be maintained by the new method, it is observed that their angle accuracy deteriorates due to the smaller $L$ value.
The angle accuracy of the microtracks is evaluated by the difference in angle when compared to the basetracks.
The width of the distribution of the angle differences is then only determined by the microtrack angle accuracy.
Figure~\ref{fig:03:microtrack_angle_accuracy} shows the angle accuracy of the microtracks recognized by using $60\,\mathrm{\mu m}$ and $30\,\mathrm{\mu m}$ thick emulsion gels.
The $30\,\mathrm{\mu m}$ case exhibits worse angle accuracy when compared to the $60\,\mathrm{\mu m}$ case.

The recognized angle of the microtrack can be expressed as $\tan\theta_{y,\mathrm{micro}} = (y_1 - y_0) / (z_1 - z_0)$ where two points, $(y_0, z_0)$ and $(y_1, z_1)$, are the end points of the track.
Using this model, the accuracy of the microtrack angle can be expressed as
\begin{equation}
    \sigma_{\tan\theta_{y,\mathrm{micro}}} = \frac{\sqrt{2}}{L}\sqrt{(\delta_y)^2+(\tan\theta_{y,\mathrm{micro}}\times\delta_z)^2}.
    \label{eq:angle_accuracy}
\end{equation}
Here, $\delta_y$ and $\delta_z$ represent the accuracies of the position recognition of the silver grains which are the parallel and perpendicular to the surface of the films, respectively.
According to equation~\eqref{eq:angle_accuracy}, a lower $L$ value reduces the angle accuracy since the angle accuracy is proportional to $1/L$.
Lower angle accuracy of microtracks results in a larger calculation time and lower S/N in the basetrack reconstruction using the microtrack connection.

The conventional method described in Refs.~\cite{Niwa:1974, Aoki:1989uk, Nakano:dt, Morishima:2010zz, Fukuda:2013nq, Yoshimoto:2017ufm, Fukuda:2014vda} was developed for relatively small angle tracks and thus, the accuracy of the microtracks for the large angle tracks is not good with the conventional microtrack recognition~\cite{Fukuda:2014vda}.

\begin{figure}[h]
    \centering 
    \includegraphics[width = 0.75\textwidth]{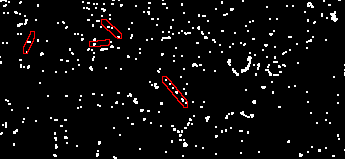}
    \caption[emulsion binary image]{A binarized image in an emulsion of a $155\,\mathrm{\mu m} \times 72\,\mathrm{\mu m}$ area. The red circles indicate the selected pixels in the linear fitting for the recognized tracks. The track in the center represents a microtrack caused by a minimum ionizing particle and the other tracks represent probably the noise contributions.\label{fig:03:microtrack_image_binary}}
\end{figure}

This method recalculate the position and angle of the track after the recognition using binarized images.
The previous analysis~\cite{Niwa:1974, Aoki:1989uk, Nakano:dt, Morishima:2010zz, Fukuda:2013nq, Yoshimoto:2017ufm, Fukuda:2014vda} shows that the recognition efficiency of the small or large angle tracks is sufficiently high, and this can be generally said to the new method.
This process involves using 16 even-numbered tomographic images.
Firstly, the pixel coordinate positions of the microtrack in the binarized images are calculated from the recognized microtrack positions and angles.
The pixels within $\pm \sqrt{\tan^2\theta_{x,\mathrm{micro}} + \tan^2\theta_{y,\mathrm{micro}}} \times 4\,\mathrm{\mu m}$ along the track and $\pm 1\,\mathrm{\mu m}$ perpendicular to the track in the image are then selected for fitting.
Figure~\ref{fig:03:microtrack_image_binary} shows the area for microtracks.

The pixels are then fitted by a linear function.
The positions and angles of the microtrack are expressed by using the least-squares method as
\begin{align}
    x &= \frac{\sum_{i=1}^{N} z_i^2 \sum_{i=1}^{N} x_i - \sum_{i=1}^{N} x_i z_i \sum_{i=1}^{N} z_i}{N \sum_{i=1}^{N} z_i^2 - (\sum_{i=1}^{N} z_i)^2} \label{eq:microtrack_leastsquares_position_x} \\
    y &= \frac{\sum_{i=1}^{N} z_i^2 \sum_{i=1}^{N} y_i - \sum_{i=1}^{N} y_i z_i \sum_{i=1}^{N} z_i}{N \sum_{i=1}^{N} z_i^2 - (\sum_{i=1}^{N} z_i)^2}. \label{eq:microtrack_leastsquares_position_y} \\
    \tan\theta_{x,\mathrm{micro}} &= \frac{N \sum_{i=1}^{N} x_i z_i - \sum_{i=1}^{N} x_i \sum_{i=1}^{N} z_i}{N \sum_{i=1}^{N} z_i^2 - (\sum_{i=1}^{N} z_i)^2} \label{eq:microtrack_leastsquares_angle_x} \\
    \tan\theta_{y,\mathrm{micro}} &= \frac{N \sum_{i=1}^{N} y_i z_i - \sum_{i=1}^{N} y_i \sum_{i=1}^{N} z_i }{N \sum_{i=1}^{N} z_i^2 - (\sum_{i=1}^{N} z_i)^2} \label{eq:microtrack_leastsquares_angle_y}
\end{align}
Here, $N$ represents the number of hit pixels, and $x_i$, $y_i$, and $z_i$ represent the positions of the hit pixels.
The above calculation is then iterated using the positions and angles in equations~\eqref{eq:microtrack_leastsquares_position_x}--\eqref{eq:microtrack_leastsquares_angle_y}.
The position and angle inputs used in the pixel selection in the iteration are substituted by the recognized positions and angles, and the pixels used in the linear fitting are reselected using the calculated values.
The values obtained by the convergence of the position and angles after several fittings are considered as the recognized positions and angles of the microtrack.

\begin{figure}[H]
    \centering 
    \includegraphics[width = 0.75\textwidth]{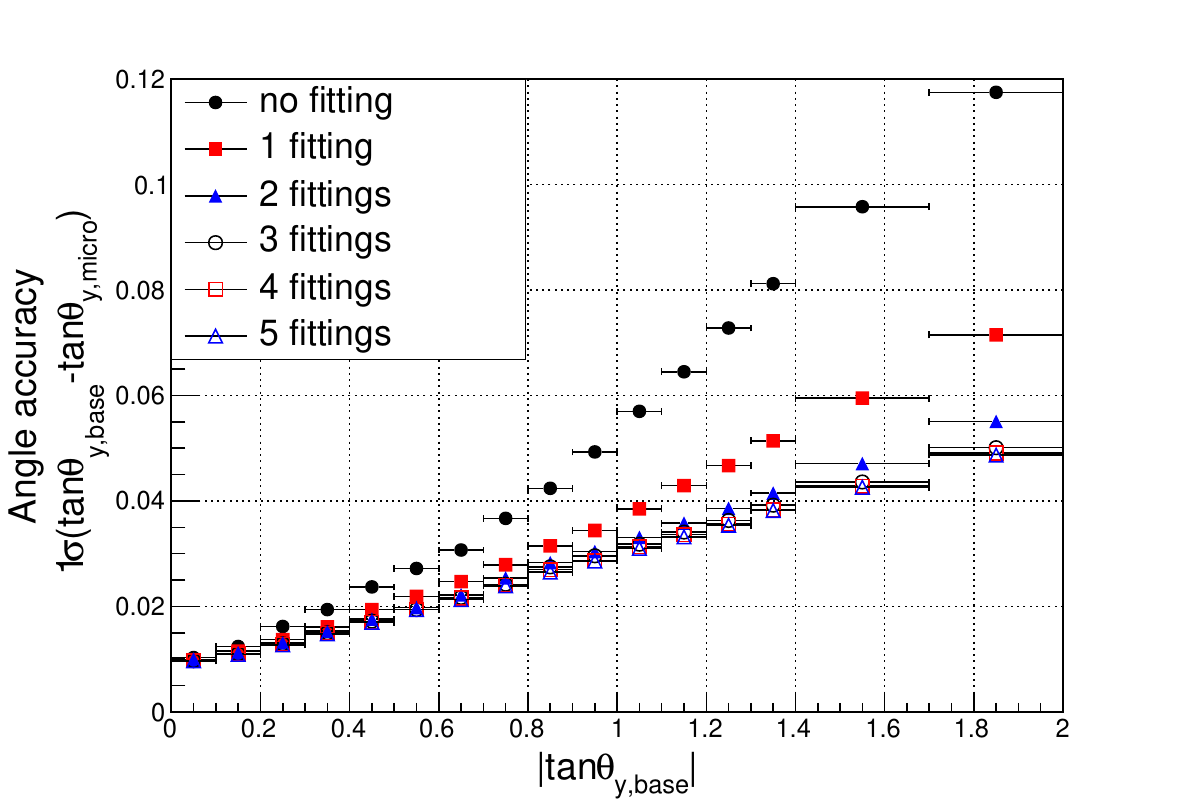}
    \caption[microtrack angle accuracy iteration thick]{The angle accuracy of the microtracks after the recalculation using the binarized images. The microtracks are detected using $60\,\mathrm{\mu m}$ thick emulsion gel layers. The colors and marker styles correspond to the number of fittings.\label{fig:03:microtrack_fitting_iteration_thick}}
\end{figure}

\begin{figure}[H]
    \centering 
    \includegraphics[width = 0.75\textwidth]{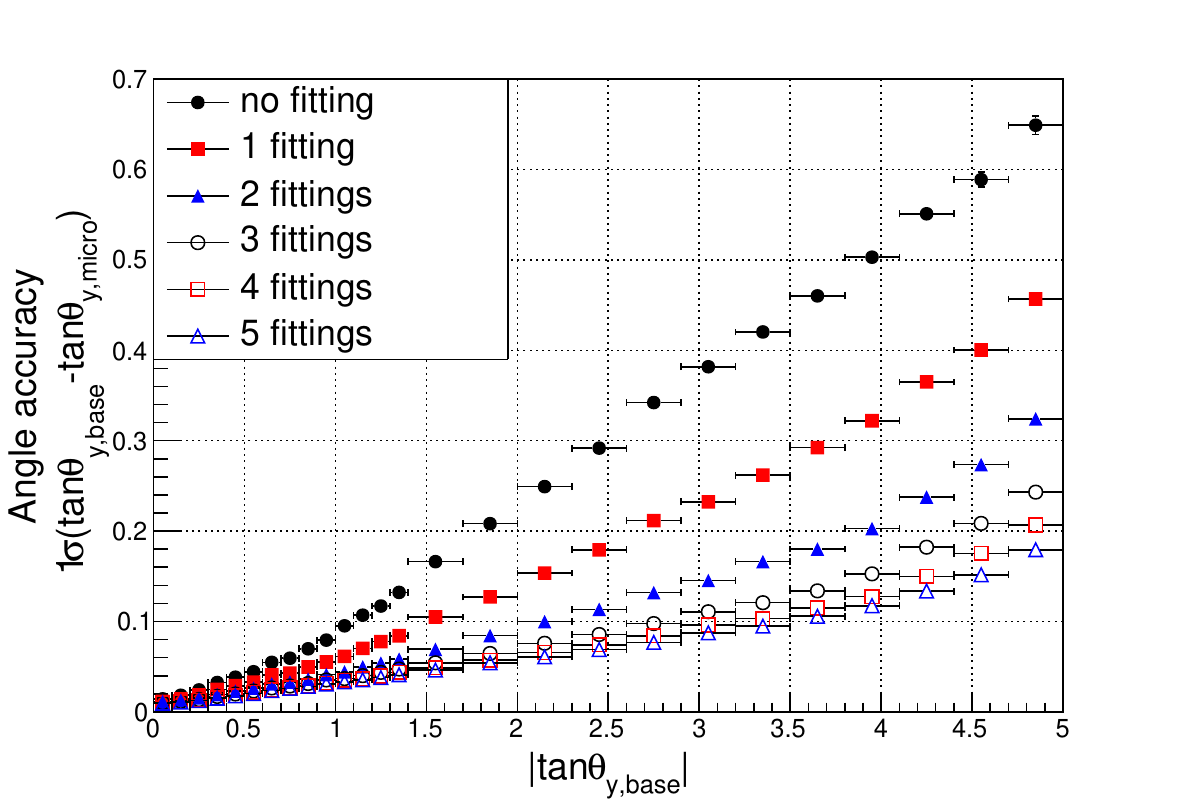}
    \caption[microtrack angle accuracy iteration thin]{The angle accuracy of the microtracks after the recalculation using the binarized images. The microtracks are detected using $30\,\mathrm{\mu m}$ thick emulsion gel layers, and the $60\,\mathrm{\mu m}$ thick binarized images are used for recalculation. The colors and marker styles correspond to the number of fittings.\label{fig:03:microtrack_fitting_iteration_thin}}
\end{figure}

Figures~\ref{fig:03:microtrack_fitting_iteration_thick} and \ref{fig:03:microtrack_fitting_iteration_thin} show the angle accuracy of the microtracks using the $60\,\mathrm{\mu m}$ and $30\,\mathrm{\mu m}$ thick emulsion gels, respectively.
The colors and marker styles of the plots represent the number of fittings.
It can be observed from these plots, that the angle accuracy converges after five iterations of fittings.
It should be noted that the angle accuracy of not only the $30\,\mathrm{\mu m}$ case but also the $60\,\mathrm{\mu m}$ case of the microtracks is improved due to this fitting.
In the conventional method of the track recognition, only the hit pixels exceeding the threshold of the track recognition is used.
On the other hand, almost all hit pixels can be used to reconstruct the track in this method as shown in Fig.~\ref{fig:03:microtrack_image_binary}.
In addition, hit pixels are searched for along the track direction in the new method, thus noise hit pixels less affect the track reconstruction.
Therefore, the angle accuracy is also improved in the $60\,\mathrm{\mu m}$ case.
Furthermore, the angle accuracy of the large angle tracks detected using the $30\,\mathrm{\mu m}$ thick emulsion gels is significantly improved.
The angle accuracy of the large angle tracks is improved by four times after the fittings.
This is because the track recognition is performed using the $30\,\mathrm{\mu m}$ thick emulsion gel, but the angle is calculated using the $60\,\mathrm{\mu m}$ thick binarized images.
The discrepancy between the small and large angle tracks shown in figure~\ref{fig:03:microtrack_angle_accuracy} is resolved by this linear fitting to the hit pixels.

\subsection{Basetrack reconstruction and noise rejection\label{ssec:analysis:basetrack}}

\begin{figure}[h]
    \centering
    \includegraphics[width = 0.8\textwidth]{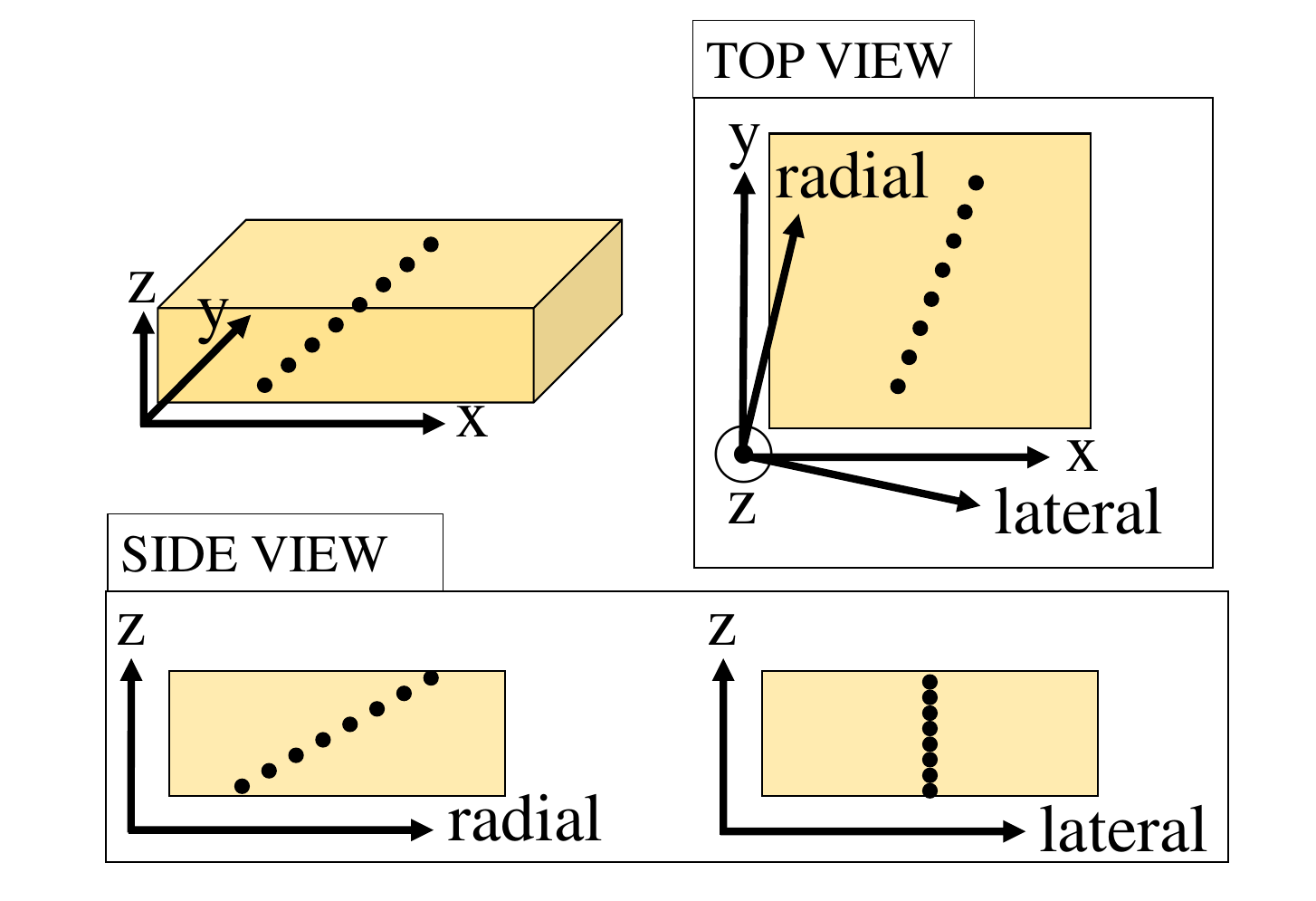}
    \caption[Definition of the radial/lateral directions.]{Definition of the radial and lateral directions. The dots indicate silver grains of the track. The lateral and radial directions are perpendicular and parallel to the track direction, respectively.\label{fig:03:radial_lateral_direction_define}}
\end{figure}

The angle of the basetrack is calculated from the two positions of the microtracks on the boundaries of the gel and base.
The number of the recognized microtracks is approximately $0.5\text{--}3 \times 10^7\,\mathrm{/cm^2}$ and that of the basetracks is approximately  $1 \times 10^5\,\mathrm{/cm^2}$.
The actual number of the tracks in the film is estimated to be only around $1\times 10^4\,\mathrm{/cm^2}$.
More than 90\% of all tracks are attributed to noise such as accidental coincidences of low energy electron tracks from environmental radioactivity or due to random noise.
The linearity and blackness of the tracks are used to remove the noise tracks~\cite{Fukuda:2010zzd, Fukuda:2017clt}.
The linearity is calculated by the differences in the angle between the basetrack and the microtracks, and the blackness is evaluated by using the Volume Pulse Height (VPH).
The VPH is a value obtained in the microtrack recognition and strongly correlates to the number of silver grains of the track~\cite{Toshito:2006xz}.
The angle accuracy of the track deteriorates for a larger track angle, as shown in equation~\eqref{eq:angle_accuracy}.
Therefore, the radial and lateral directions in figure~\ref{fig:03:radial_lateral_direction_define} are used in the evaluation of the linearity.
The lateral and radial directions are perpendicular and parallel to the track direction, respectively.
A track can be considered as the perpendicular track in the $z$-lateral plane and the angle accuracy does not change in principle, as shown in figure~\ref{fig:03:radial_lateral_direction_define}.
However, the accuracy deteriorates when the track angle increases along the radial direction.

\begin{figure}[H]
    \centering
    \includegraphics[width = 0.8\textwidth]{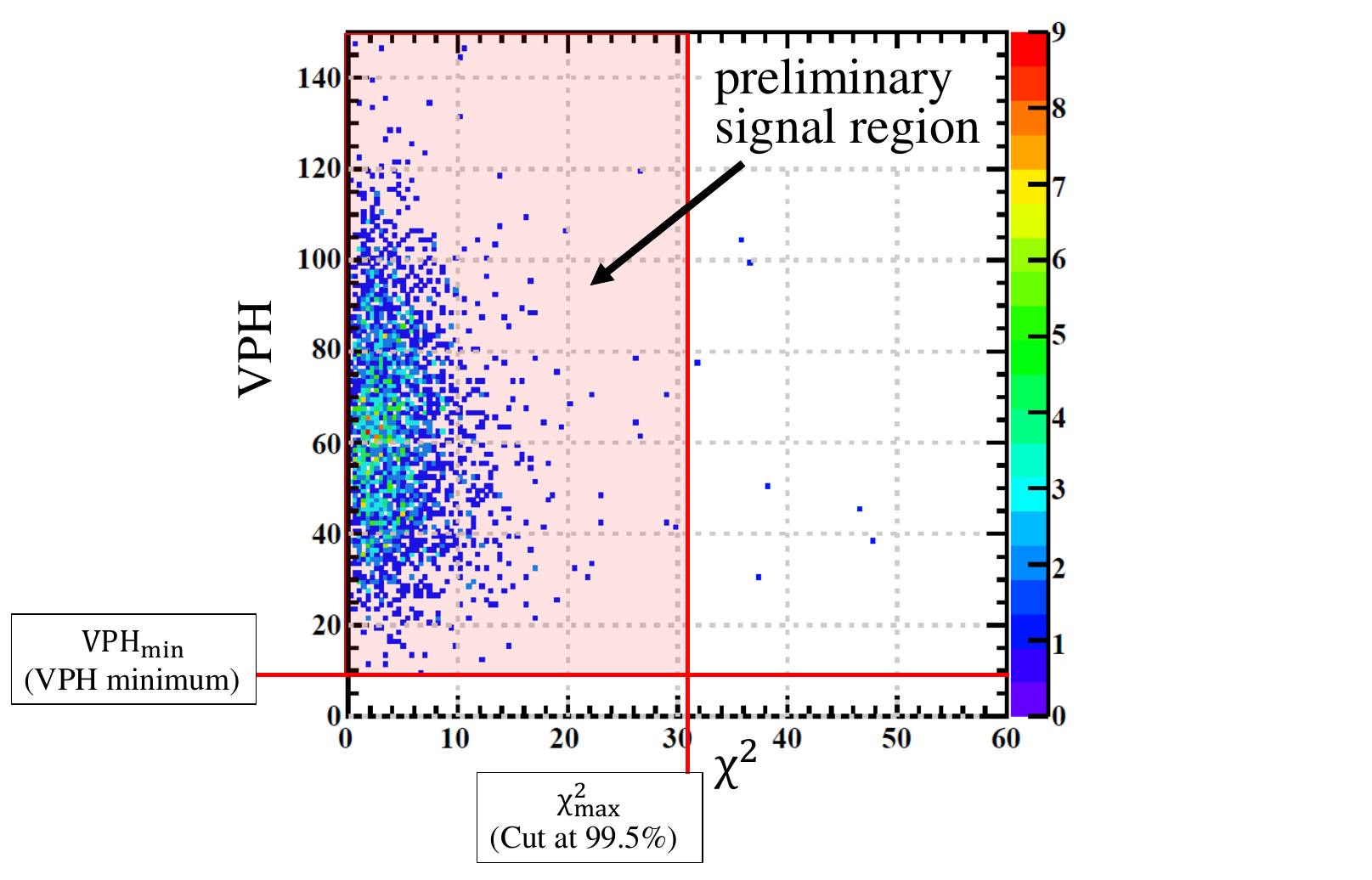}
    \caption[connected basetrack adjacent films]{The two-dimensional distribution of the basetracks connected to the adjacent films. The horizontal and vertical axes represent the linearity and blackness of the tracks, respectively. The area within the red lines is defined as the preliminary signal region. Here, the $\chi^2$ line is determined to contain 99.5\% of the distribution, and that of VPH is determined to contain all the VPH limited by the scanning.\label{fig:03:basetrack_connected_adjacent_films}}
\end{figure}

Four angle differences can be obtained along the radial and lateral directions with respect to the basetrack direction.
Using these angle differences, the linearity of the track is defined as
\begin{equation}
\chi^2 = \left(\frac{d\tan\theta_{0,r}}{\sigma_\mathrm{radial}}\right)^2 + \left(\frac{d\tan\theta_{0,l}}{\sigma_\mathrm{lateral}}\right)^2 + \left(\frac{d\tan\theta_{1,r}}{\sigma_\mathrm{radial}}\right)^2 + \left(\frac{d\tan\theta_{1,l}}{\sigma_\mathrm{lateral}}\right)^2
\label{eq:track_linearity}
\end{equation}
where, $\sigma_\mathrm{radial}$ represents the angle accuracy shown in figure~\ref{fig:03:microtrack_fitting_iteration_thin}, and $\sigma_\mathrm{lateral}$ is the angle accuracy with $\tan\theta_{x(y),\mathrm{base}} = 0$.
$d \tan\theta_{i,p}$ is the angle difference between the microtracks and basetrack.
Here, the subscripts $i = 0$ and $1$ represent the angle difference between the upper and lower microtracks, respectively, and $p = r$ and $l$ represent the angle difference in the radial and lateral views, respectively.
Since the angle differences of the signal track, which penetrates the film, are expected to follow the Gauss distribution, the linearity follows the chi-square distribution of four degrees of freedom.
Figure~\ref{fig:03:basetrack_connected_adjacent_films} shows a two-dimensional distribution of the linearity and the blackness of the basetracks connected to adjacent films.
The requirement of the connection to adjacent films extracts tracks penetrating the film.
When there are more noise, it is required that the tracks are connected to not only adjacent films but also multiple films, or tighter connection allowance is applied to increase the reliability of the track.
When the tracks are connected between adjacent films, the position differences of correctly connected tracks show the Gaussian distribution while the mis-connection due to the chance coincidence is uniformly distributed.
A contamination from the accidental coincidence can be estimated from the distribution and it was 0.5\% in this analysis.
Such a contamination is expected to have larger $\chi^2$ of linearity. 
Therefore, the preliminary signal region is defined as the area enclosed within the red lines in figure~\ref{fig:03:basetrack_connected_adjacent_films}.
Here, it is determined that the line of $\chi^2$ contains 99.5\% of the distribution, and that of VPH contains all the VPH allowed by the scanning.
Hereafter, the maximum value of $\chi^2$ of the preliminary signal region is denoted as $\chi^2_\mathrm{max}$ and the minimum value and median of the VPH are denoted as $\mathrm{VPH}_\mathrm{min}$ and $\mathrm{VPH}_\mathrm{med}$, respectively.

\begin{figure}[H]
    \centering 
    \includegraphics[width = 0.475\textwidth]{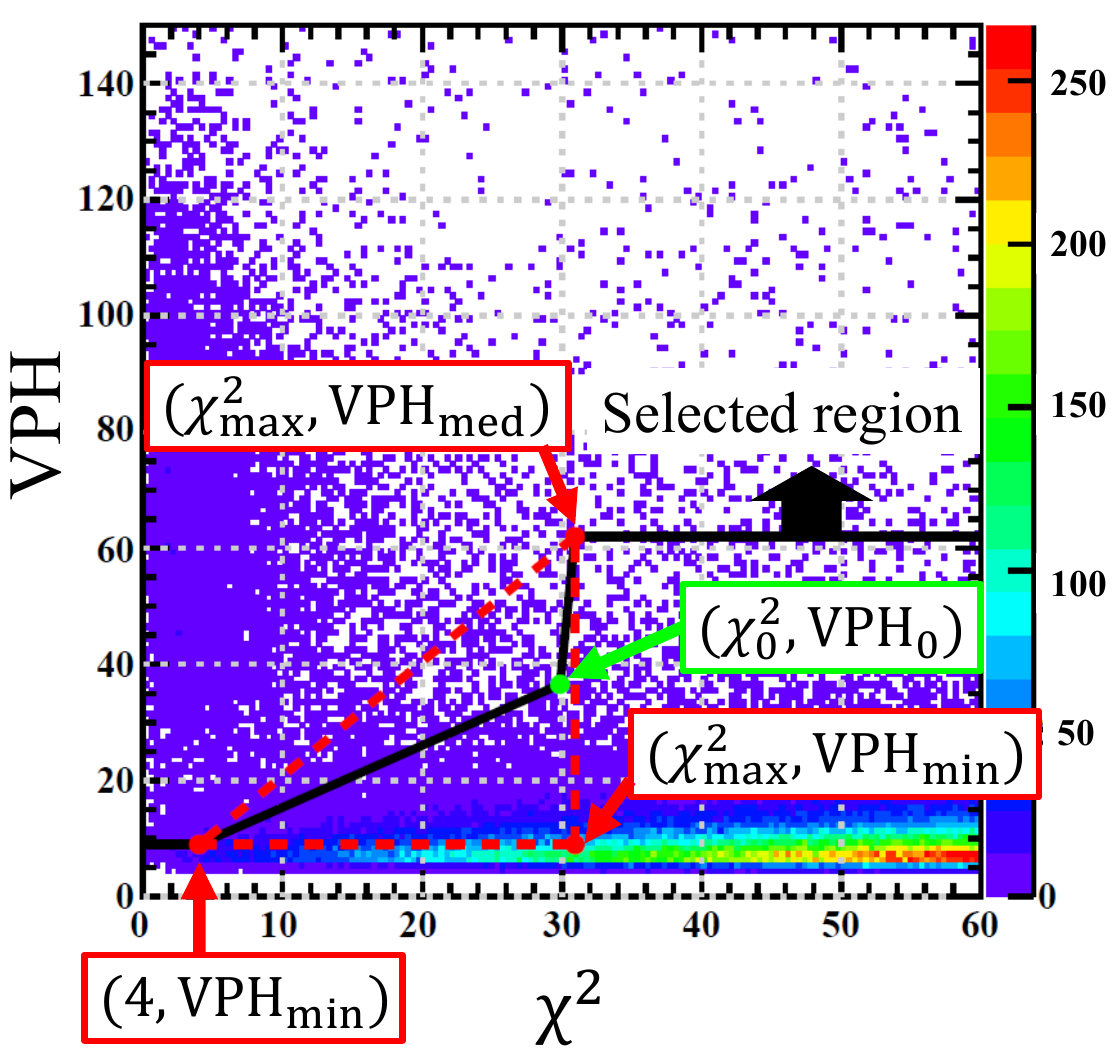}
    \includegraphics[width = 0.475\textwidth]{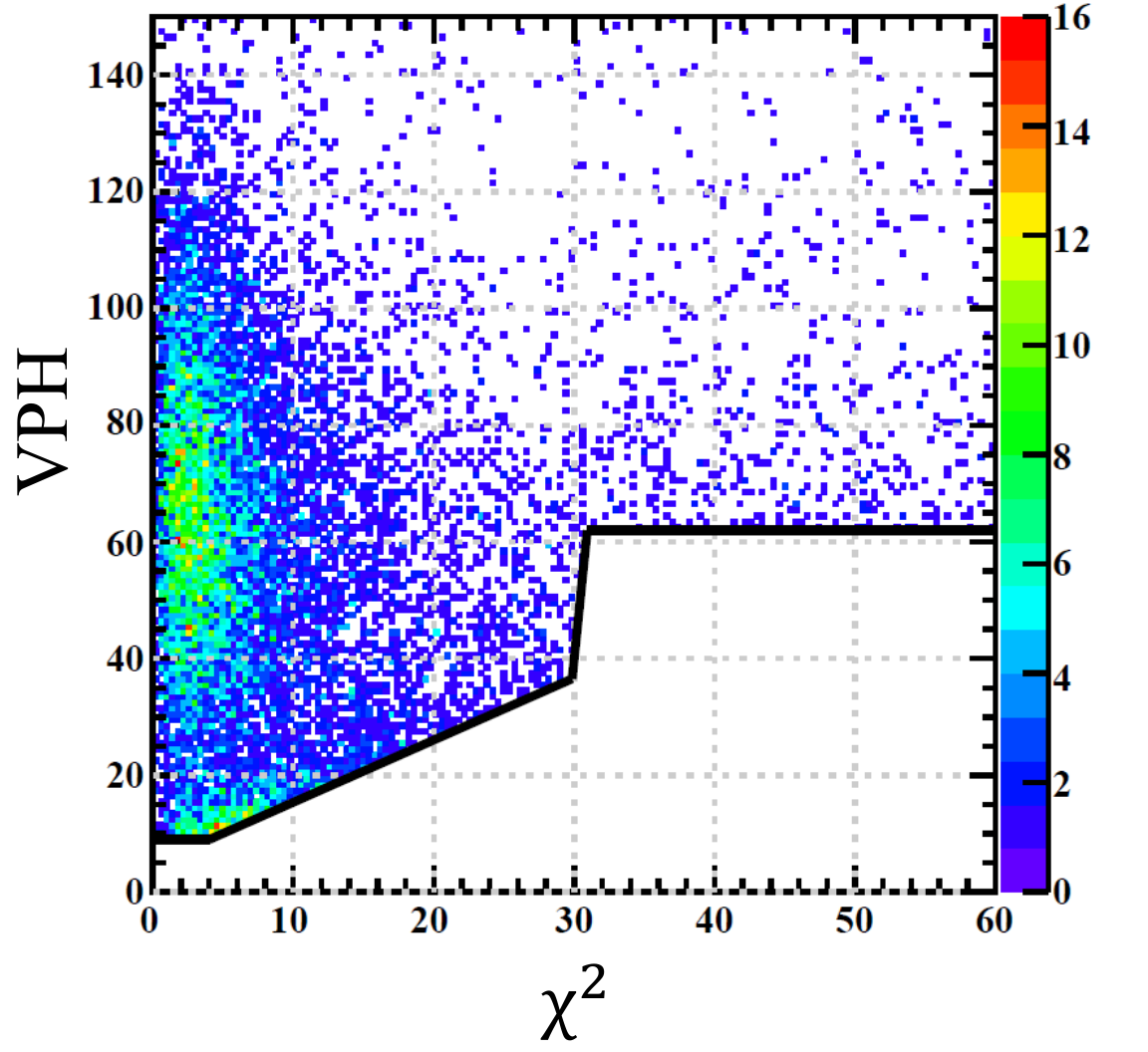}
    \caption[basetrack ranking cut]{The two-dimensional distribution of the basetracks. The horizontal axis represents the linearity of the track, $\chi^2$ and the vertical axis represents the blackness of the track, VPH. The left figure shows all basetracks and the black line represents the selection cut. The noise tracks have larger $\chi^2$ of the linearity and small VPH. The noise contamination below the selection cut black lines is removed. The selected basetracks are shown in the right figure. The cut criteria are explained in the last paragraph of Sect.~\ref{ssec:analysis:basetrack}.\label{fig:03:basetrack_ranking_cut}}
\end{figure}

Figure~\ref{fig:03:basetrack_ranking_cut} presents all the basetracks in a single emulsion film including those not connected to the adjacent films.
The contribution of the fake tracks can be clearly observed in the region where $\chi^2$ is large and $\mathrm{VPH}$ is small.
The cut is applied as shown in the black lines in figure~\ref{fig:03:basetrack_ranking_cut} to reduce such a noise.
The cut consists of four lines, and is expressed as
\begin{equation}
\mathrm{VPH} \geq \begin{dcases}
    \mathrm{VPH_{min}} & (0 \leq \chi^2 < 4) \\
    \frac{\mathrm{VPH}_0 - \mathrm{VPH_{min}}}{\chi^2_0 - 4}(\chi^2 - 4) + \mathrm{VPH_{min}} & (4 \leq \chi^2 < \chi^2_0) \\
    \frac{\mathrm{VPH_{med}} - \mathrm{VPH}_0}{\chi^2_\mathrm{max} - \chi^2_0}(\chi^2 - \chi^2_0) + \mathrm{VPH_0} & (\chi^2_0 \leq \chi^2 < \chi^2_\mathrm{max}) \\
    \mathrm{VPH_{med}} & (\chi^2_\mathrm{max} \leq \chi^2).
\end{dcases}
\label{eq:basetrack_ranking_condition}
\end{equation}
Here, 4 represents the mean of the $\chi^2$ distribution with four degrees of freedom.
The point, $(\chi^2_0, \mathrm{VPH}_0)$, is defined as follows.
Firstly, a triangle is formed by the points, $(4, \mathrm{VPH}_\mathrm{min})$, $(\chi^2_\mathrm{max}, \mathrm{VPH}_\mathrm{min})$, and $(\chi^2_\mathrm{max}, \mathrm{VPH}_\mathrm{med})$.
The point is determined to ensure that the number of tracks is the smallest while ensuring that more than 99.8\% of the tracks remain in the preliminary signal region.
The point is searched for from the lattice points in $[4, \chi^2_\mathrm{max})$ divided into 100 and $[\mathrm{VPH}_\mathrm{min}, \mathrm{VPH}_\mathrm{max})$ divided into 50 in the triangle.
The main component of the noise can be removed while maintaining a sufficient recognition efficiency of the track penetrating the film due to this cut.
Remaining component of the noise can be also removed when the tracks are connected between adjacent films and the analysis does not deteriorate with them.
\section{Performance of the large-angle track recognition\label{sec:performance}}

\subsection{Angle accuracy\label{ssec:performance:angle_precision}}

\begin{figure}[h]
    \centering 
    \includegraphics[width = 0.75\textwidth]{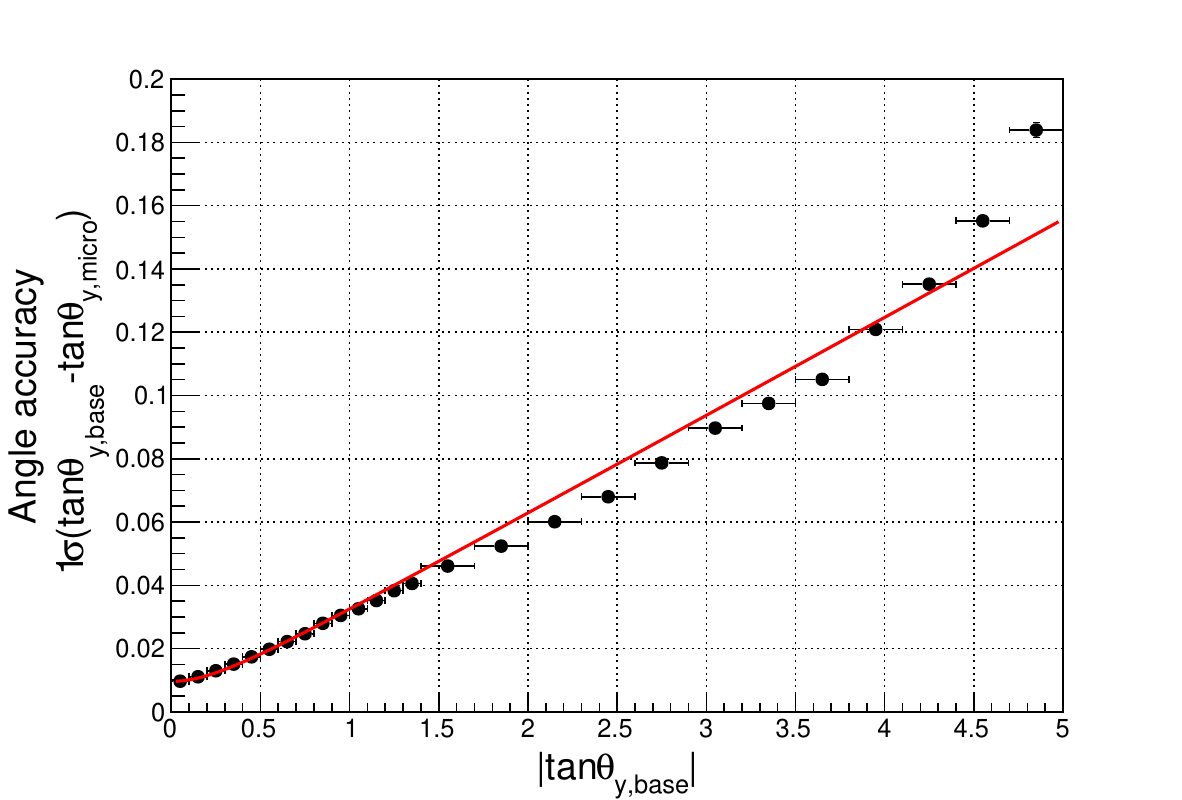}
    \caption[microtrack angle accuracy fit]{The angle accuracy of the microtracks. The plot is fitted by the function expressed in equation~\eqref{eq:angle_accuracy} and the result is $\delta_y = 0.41 \pm 0.01\,\mathrm{\mu m}$ and $\delta_z = 1.32 \pm 0.02\,\mathrm{\mu m}$.\label{fig:04:microtrack_fitting_iteration_angle_acc_fit}}
\end{figure}

The angle accuracy of the microtracks is evaluated by using the angle differences from the basetracks, as explained in Sect.~\ref{ssec:analysis:microtrack_new}.
Figure~\ref{fig:04:microtrack_fitting_iteration_angle_acc_fit} shows the angle accuracy of the microtracks recognized by the $60\,\mathrm{\mu m}$ and $30\,\mathrm{\mu m}$ thick emulsion gel layers.
The discrepancy of the angle accuracy observed when comparing figures ~\ref{fig:03:microtrack_angle_accuracy} and \ref{fig:04:microtrack_fitting_iteration_angle_acc_fit}, has been reduced significantly due to the linear fitting to the binarized images.
This angle accuracy plot is fitted by using equation~\eqref{eq:angle_accuracy} and the result is $\delta_y = 0.41 \pm 0.01\,\mathrm{\mu m}$ and $\delta_z = 1.32 \pm 0.02\,\mathrm{\mu m}$.
$\delta_z$ is better than the depth of field of HTS, $4\,\mathrm{\mu m}$.
This is because the measurement of the angle uses not only two position information but that of all the hit pixels in the 16 tomographic images.
When the angle of the track gets larger, it tends not to be fully detected by one sensor.
In this case, the angle of the track is reconstructed with less information and the angle accuracy gets worse.
Therefore, data points and fitted curve is inconsistent in the area of $|\tan\theta_{y,\mathrm{base}}| > 4.5$ in Fig.~\ref{fig:04:microtrack_fitting_iteration_angle_acc_fit}.

\begin{figure}[h]
    \centering
    \includegraphics[width = 0.65\textwidth]{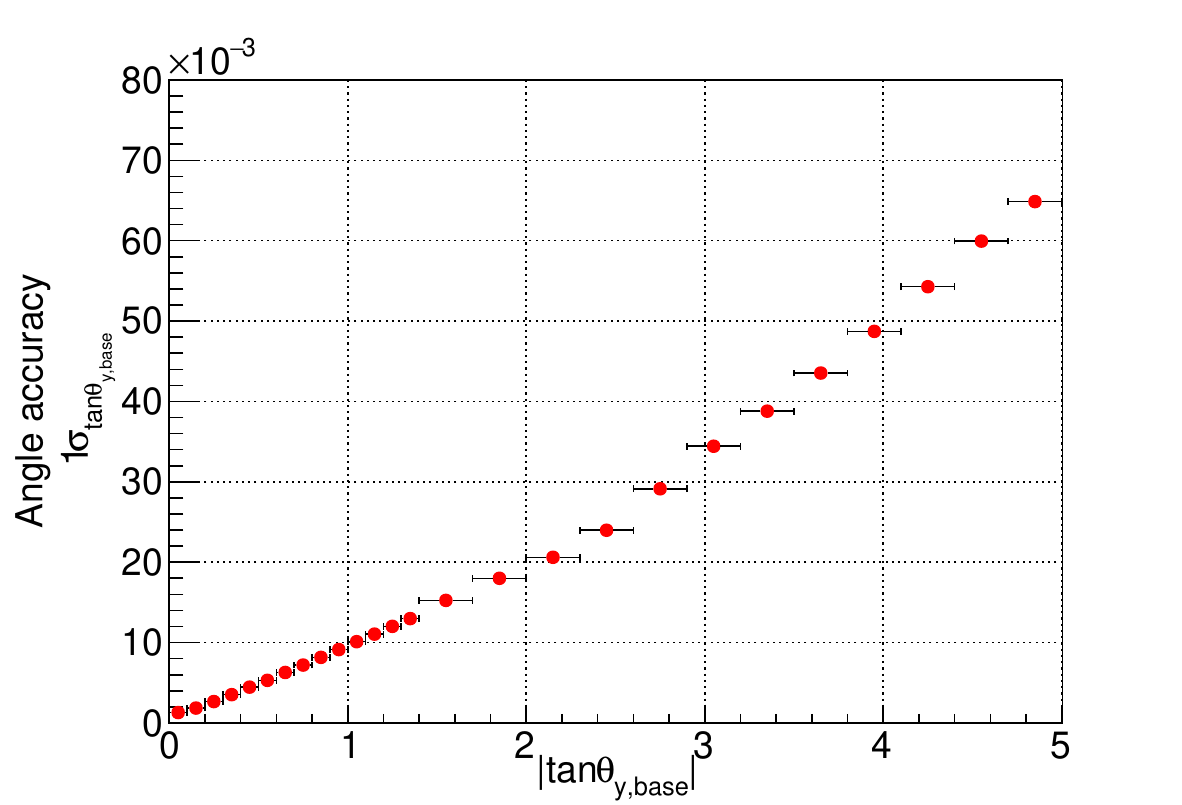}
    \caption[basetrack angle accuracy fit]{Angle accuracy of the basetracks. The basetracks can be measured with an accuracy of 0.065 even for large angle tracks ($|\tan\theta_{x(y),\mathrm{base}}| \simeq 5.0$).\label{fig:04:basetrack_precision}}
\end{figure}

The angle accuracy of the basetracks is then evaluated using the angle difference between two adjacent films in the ECC.
There are no materials between the films used in this evaluation.
Thus, the angle difference ($\Delta\tan\theta_{x(y),\mathrm{base}}$) is obtained from those films and the angle accuracy is evaluated.
The basetracks can be measured with an accuracy of 0.065 even for large angle tracks ($|\tan\theta_{x(y),\mathrm{base}}| \simeq 5.0$), as shown in figure~\ref{fig:04:basetrack_precision}.

\begin{figure}[h]
    \centering
    \includegraphics[width = 0.65\textwidth]{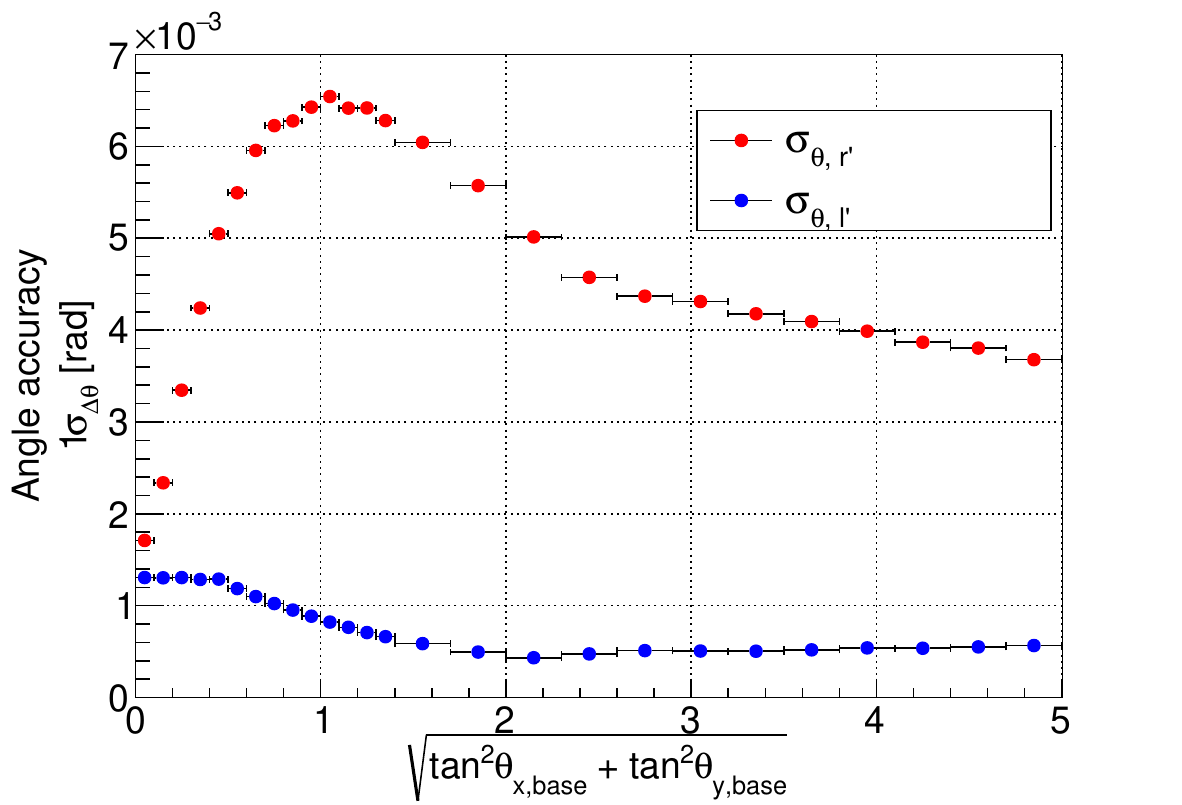}
    \caption[new radial lateral accuracy]{Widths of Gaussian function fitted to the angle differences. Here, the angle difference represents the difference of $\theta$ and not that of $\tan\theta$. $\sigma_{\theta,\mathrm{l'}}$, plotted with blue points, represents a width of the distribution of the angle differences projected to the $\hat{t}$-$\hat{l}'$ plane, and $\sigma_{\theta,\mathrm{r'}}$, plotted with red points, represents that projected to the $\hat{t}$-$\hat{r}'$ plane.\label{fig:04:new_rad_lat_precision}}
\end{figure}

\begin{figure}[h]
    \centering
    \includegraphics[width = 0.4\textwidth]{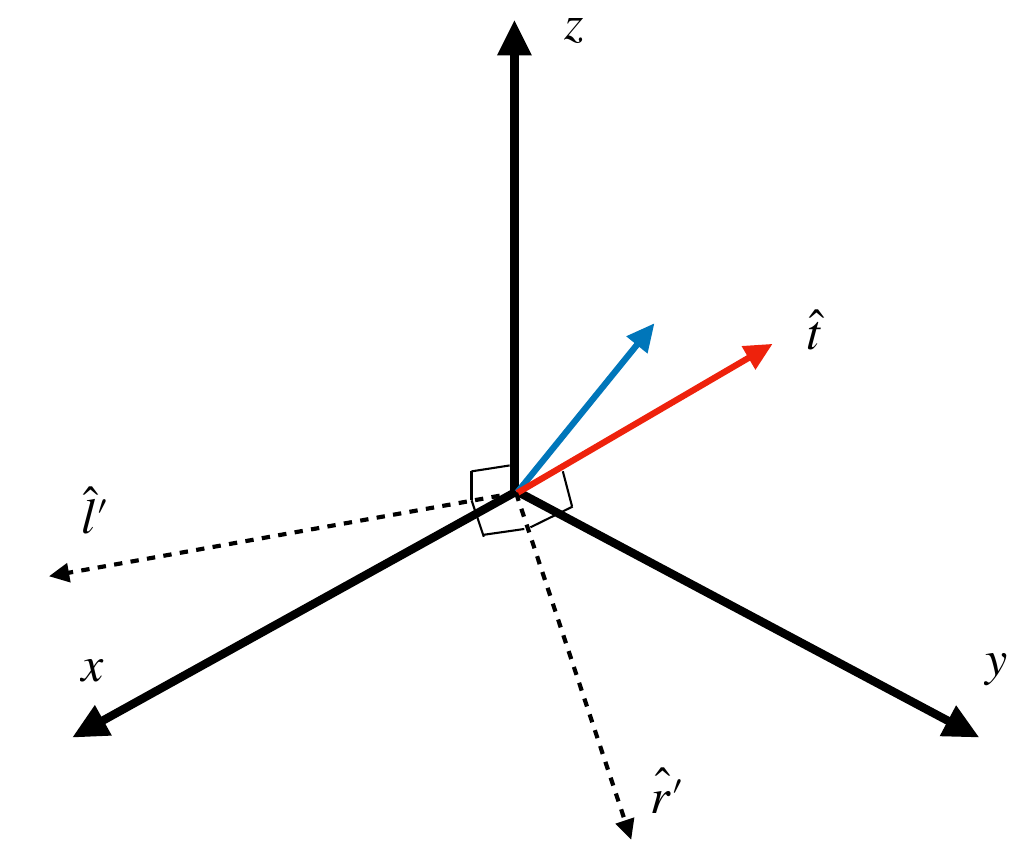}
    \caption[new radial lateral directions]{The angle difference is projected on two planes formed by the directions, $\hat{t}$, $\hat{l}'$, and $\hat{r}'$. $\hat{t}$ is the upstream track direction, $\hat{l}' = \hat{t} \times \hat{z}$, and $\hat{r}' = \hat{t} \times \hat{l}'$. \label{fig:04:mcs_coordinate}}
\end{figure}

Figure~\ref{fig:04:new_rad_lat_precision} shows the width of the distributions of the angle differences of two adjacent basetracks.
Here, the angle difference represents the difference of $\theta$ and not that of $\tan\theta$.
The angle difference is projected on two planes formed by the directions defined below.
$\hat{t}$ is the upstream track direction, and $\hat{l}'$ and $\hat{r}'$ are defined as
\begin{align}
    \hat{l}' &= \hat{t} \times \hat{z} \\
    \hat{r}' &= \hat{t} \times \hat{l}'
\end{align}
The two angle differences are measured on the $\hat{t}$-$\hat{l}'$ and $\hat{t}$-$\hat{r}'$ planes.
$\sigma_{\theta,\mathrm{l'}}$ and $\sigma_{\theta,\mathrm{r'}}$ are the widths of the Gaussian function fitted to the angle differences on the $\hat{t}$-$\hat{l}'$ and $\hat{t}$-$\hat{r}'$ planes, respectively.
When $\hat{t} \parallel \hat{z}$, the $\hat{z}$-$\hat{x}$ and $\hat{z}$-$\hat{y}$ planes are defined as the $\hat{t}$-$\hat{l}'$ and $\hat{t}$-$\hat{l}'$ planes, respectively.
Figure~\ref{fig:04:mcs_coordinate} shows the directions, $\hat{r}'$ and $\hat{l}'$.
The directions $\hat{r}'$ and $\hat{l}'$ are shown in figure~\ref{fig:04:mcs_coordinate}.
These angle accuracies enable momentum reconstruction even for large angles using multiple Coulomb scattering~\cite{Hiramoto:2020gup, Oshima:2020ozn}.
Particularly, $\sigma_{\theta,\mathrm{l'}}$ is $\sim 1.5\,\mathrm{mrad}$ and can be sufficiently used for the reconstruction of momentum up to $\mathcal{O}(\mathrm{GeV}/c)$.

\subsection{Basetrack reconstruction efficiency and noise rejection\label{ssec:analysis:deteff_sn}}

Two microtracks can be obtained in the $60\,\mathrm{\mu m}$-thickness scanning, and two microtracks can also be obtained in the inner and outer half of the emulsion gel layer by the $30\,\mathrm{\mu m}$-thickness scanning in one emulsion gel layer, as shown in figure~\ref{fig:03:film_crosssection_32scan}.
Therefore, the basetracks can be classified into four types in each scanning thickness based on the microtracks that it comprises.
Each type of the basetracks are named as shown in table~\ref{tab:04:microtrack_definition}.
\begin{table}[t]
    \centering
    \caption[basetrack type]{Basetrack names and the microtracks consisting of them. Four types of basetracks are reconstructed from $2\times2$ microtracks.\label{tab:04:microtrack_definition}}
    \begin{tabular}{ccccc} \hline
        {} & \multicolumn{2}{c}{$60\,\mathrm{\mu m}$-thickness scanning} & \multicolumn{2}{c}{$30\,\mathrm{\mu m}$-thickness scanning} \\
        {} & Upper & Down & Upper & Down \\ \hline
        basetrack 0 & Even & Even & Inner & Inner \\
        basetrack 1 & Even & Odd  & Inner & Outer \\
        basetrack 2 & Odd  & Even & Outer & Inner \\
        basetrack 3 & Odd  & Odd  & Outer & Outer \\ \hline
    \end{tabular}
\end{table}
Additionally, the below selections are also considered to increase the basetrack reconstruction efficiency.
\begin{itemize}
    \item basetrack ${}_4\mathrm{C}_3$ : at least three out of $2 \times 2$ microtracks are recognized.
    \item basetrack all : at least one microtrack in the upstream emulsion gel layer and one microtrack in the downstream emulsion gel layer are recognized.
\end{itemize}
The numbers of reconstructed tracks and their reconstruction efficiencies are evaluated in one film in the ECC for these types of basetracks. The tracks used for this evaluation are mainly due to muons passed through it and they are selected based on the conditions described below.
\begin{enumerate}
    \item Tracks consisting of corresponding basetracks in 4 plates, two in upstream and two in downstream of the film to be evaluated, are used.
    \item Tracks penetrating five or more iron plates are selected from the tracks.
    \item Further criteria to limit high energy muons, the standard deviation of angle differences across one iron plate projected on the $\hat{t}$-$\hat{l}'$ plane is less than 2\,mrad, are applied.
\end{enumerate}
The basetrack in the downstream film is extrapolated to the evaluated film after these cuts.

\begin{figure}[h]
    \centering 
    \includegraphics[width = 0.75\textwidth]{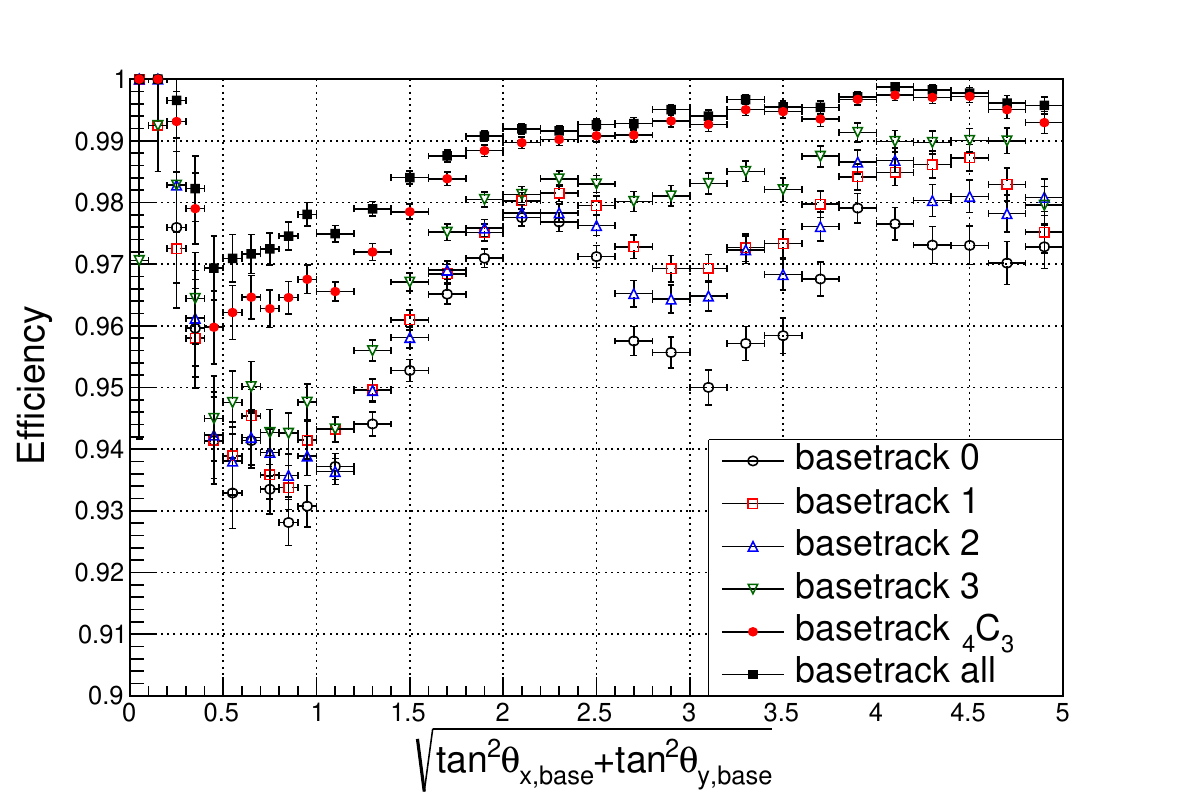}
    \caption[basetrack efficiency]{The angle dependence of the basetrack reconstruction efficiency. The colors and marker styles correspond to the combinations of the microtracks used to reconstruct the basetrack.\label{fig:04:basetrack_efficiency}}
\end{figure}

\begin{figure}[h]
    \centering 
    \includegraphics[width = 0.75\textwidth]{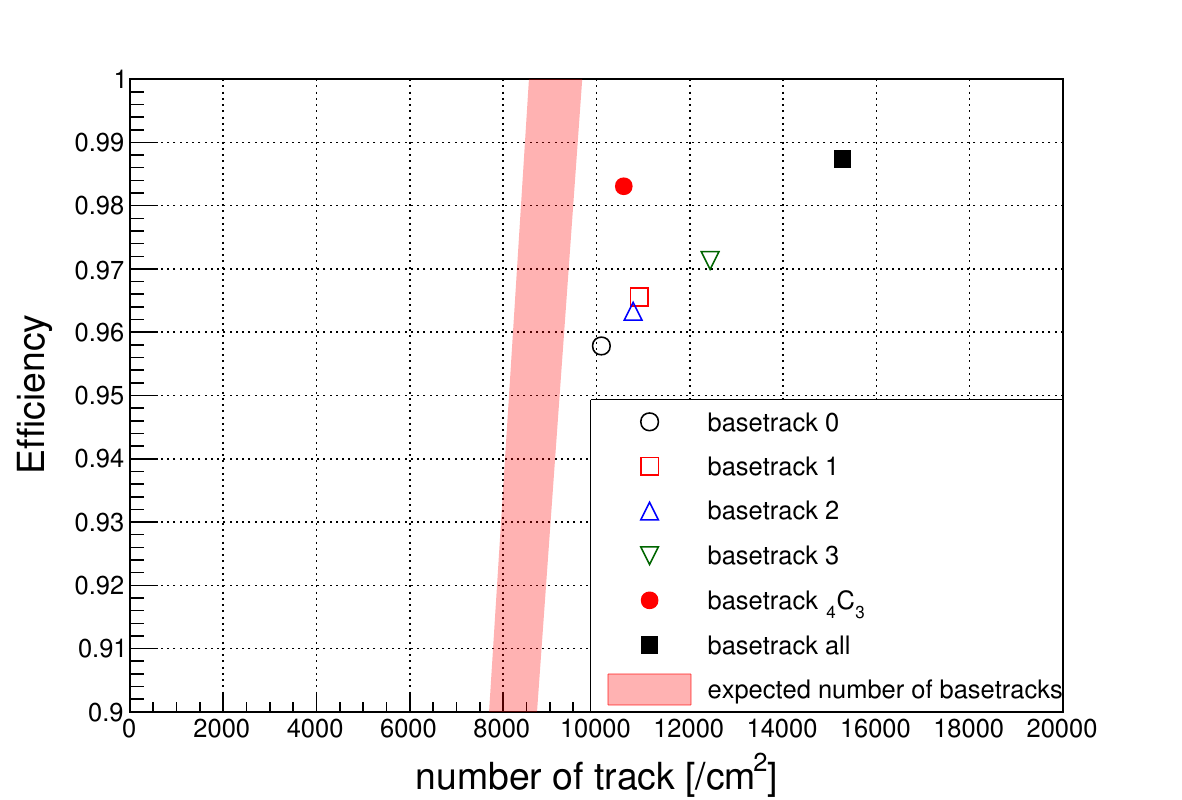}
    \caption[basetrack efficiency number of track]{The basetrack reconstruction efficiency and the number of tracks ($|\tan\theta_{x(y),\mathrm{base}}| < 5.0$). The colors and marker styles correspond to the combinations of the microtracks to reconstruct the basetrack. The red band indicates the expected number of basetracks estimated from the number of all the cosmic ray muons accumulated until the development. The slope of the band corresponds to the reconstruction efficiency.\label{fig:04:basetrack_efficiency_tracknum}}
\end{figure}

The reconstruction efficiency is defined as a ratio of the number of tracks with the same positions and angles as the prediction to the total number of prediction tracks.
Figure~\ref{fig:04:basetrack_efficiency} shows the reconstruction efficiency.
Figure~\ref{fig:04:basetrack_efficiency_tracknum} presents the averaged reconstruction efficiency over all angles and the number of reconstructed basetracks.
The red band indicates the estimated number of recorded tracks until the film development.
The main component of the tracks are accumulated cosmic ray muons and can be estimated by the number of days.
Noise contribution in the basetrack ${}_4\mathrm{C}_3$ selection is approximately 1,600 tracks\,$/\mathrm{cm^2}$, while in the basetrack all selection is 6,300 tracks\,$/\mathrm{cm^2}$.
The reconstruction efficiency is improved when two microtracks in the upstream (downstream) emulsion gel layers are used to reconstruct the basetrack.
Additionally, the basetrack ${}_4\mathrm{C}_3$, and basetrack all have similar reconstruction efficiencies but the number of tracks is less for the basetrack ${}_4\mathrm{C}_3$, and it has better S/N.
This is because the basetrack ${}_4\mathrm{C}_3$, requires the same track and is detected twice in at least one emulsion gel layer.
Thus, irreproducible noises are reduced in the case of basetrack ${}_4\mathrm{C}_3$.
The number of tracks is similar to the conventional track reconstruction with 16 tomographic images while the efficiency increases by 2\% and the averaged efficiency is 98\% because of the basetrack ${}_4\mathrm{C}_3$ selection.
The number of reconstructed tracks is 70\% of the cases where the basetrack is reconstructed with at least one microtrack in each emulsion gel. 

\newpage

\subsection{Processing speed\label{ssec:performance:process_speed}}

Table~\ref{tab:04:process_speed_summary} summarizes the throughput of each process in the new method.
\begin{table}[t]
    \centering
    \caption[process speed summary]{Summary of the throughput of the processes\label{tab:04:process_speed_summary}}
    \begin{tabular}{ccc} \hline
    Process & Number of processor & Throughput ($\mathrm{cm^2/h}$) \\ \hline
    Scan & HTS (computers $\times$ 36)~\cite{Yoshimoto:2017ufm} & 630 \\
    Microtrack recognition & Computers $\times$ 9 & 170 \\
    Basetrack reconstruction & Computers $\times$ 2 & 160 \\ \hline
    \end{tabular}
\end{table}

Although there is still room for improvement in the image encoding algorithm or in the CPU power, the throughput for the image capturing, binarization, and storing of images by HTS is measured to be $630\,\mathrm{cm^2/h}$ in this study.

The throughput for the microtrack recognition and linear fitting using the binarized images is $170\,\mathrm{cm^2/h}$.
In this process, microtracks are recognized using the combinations of the 16 binarized images described in figure~\ref{fig:03:film_crosssection_32scan}.
Linear fitting is then applied to all the recognized tracks in 16 even-numbered binarized images.
The microtrack recognition is performed on GPU and the linear fitting is performed on CPU.
Nine computers are used in this process\footnote{CPU: Intel\textregistered~Core\texttrademark~i9-10980XE, GPU: GeForce RTX\texttrademark~2080 SUPER$\times 2$ for seven computers and GeForce RTX\texttrademark~3090 for two ones}.

The throughput for the basetrack reconstruction and noise rejection is $160\,\mathrm{cm^2/h}$.
Four processes are run by two computers asynchronously\footnote{CPU : AMD Ryzen\texttrademark~Threadripper\texttrademark~2950X and AMD Ryzen\texttrademark~9~3950X}.
Due to the improvement of the angle accuracy of microtracks, the searching area for the track connection is small and the processing speed is ten times faster than that of the no-fitting case.
The track reconstruction is implemented within $|\tan\theta| < 5$ with a speed of $150\,\mathrm{m^2/year}$.
 
\section{Conclusion\label{sec:conclusion}}

This paper presents a new method of emulsion film scanning with a wide angle acceptance and practical speed, and also demonstrates its performance.
The performance of the conventional method of the emulsion film scanning by HTS specializing in the scanning speed has been evaluated only for $|\tan\theta_{x(y),\mathrm{micro}}| < 1.5$.
In the new method, HTS captures thinner tomographic images when compared to the conventional method and enables the measurement of larger angle tracks.
The linear fitting in the track recognition process is iterated to the binarized tomographic images and the accuracy of the track angles has been improved.
The angle accuracy of microtrack for $|\tan\theta_{x(y),\mathrm{micro}}| > 2.0$ is improved by four times due to linear fitting.
The method implements the track reconstruction speed around $150\,\mathrm{m^2/year}$ with the angle acceptance of $|\tan\theta_{x(y),\mathrm{base}}| < 5.0$.
The track reconstruction efficiency is also maintained higher than 98\% due to the basetrack ${}_4\mathrm{C}_3$ selection, while the number of reconstructed tracks is 70\% of the cases where the basetrack is reconstructed with at least one microtrack in each emulsion gel.
The selection reduces the irreproducible noise tracks, and the tracks from charged particles penetrating an emulsion film remain with this selection.
Therefore, both the efficiency and S/N of the new method are higher than the conventional method.
This method is essential for the scanning of a large number of films with a wide angle acceptance for the charged particles.
The total area of the emulsion films used in the J-PARC E71a of the NINJA experiment is $9.3 \times 10^5\,\mathrm{cm^2}$, due to which all the films can be processed in approximately 250\,days.
 
\section*{Acknowledgment}

We thank the NINJA collaboration for the data acquisition and fruitful discussion.
This work is supported by the JST-SENTAN Program from the Japan Science and Technology Agency, JST, and MEXT and JSPS KAKENHI Grant Numbers JP17H02888, JP18H03701, JP18H05210, JP18H05535, JP18H05537, JP18H05541, JP20J15496, JP20J20304, and JP21H01108. 
\newpage

\bibliographystyle{ptephy}

\end{document}